\documentclass[journal=ancac3,manuscript=article]{achemso}

\usepackage{lmodern}
\usepackage[T1]{fontenc}

\usepackage{bm}
\usepackage{amsmath}
\usepackage{amssymb}
\usepackage{xcolor}

\usepackage{epstopdf}
\author{Weihua~Wang}
\affiliation[Department of Photonics Engineering, Technical University of Denmark, DK-2800 Kgs. Lyngby, Denmark]{DTU Fotonik}
\alsoaffiliation[Center for Nanostructured Graphene, Technical University of Denmark, DK-2800 Kgs. Lyngby, Denmark]{Center for Nanostructured Graphene}
\author{Thomas~Christensen}
\affiliation[Department of Photonics Engineering, Technical University of Denmark, DK-2800 Kgs. Lyngby, Denmark]{DTU Fotonik}
\alsoaffiliation[Center for Nanostructured Graphene, Technical University of Denmark, DK-2800 Kgs. Lyngby, Denmark]{Center for Nanostructured Graphene}
\author{Antti-Pekka~Jauho}
\affiliation[Department of Micro- and Nanotechnology, Technical University of Denmark, DK-2800 Kgs. Lyngby, Denmark]{DTU Nanotech}
\alsoaffiliation[Center for Nanostructured Graphene, Technical University of Denmark, DK-2800 Kgs. Lyngby, Denmark]{Center for Nanostructured Graphene}
\author{Kristian~S.~Thygesen}
\affiliation[Department of Physics, Technical University of Denmark, DK-2800 Kgs. Lyngby, Denmark]{DTU Physics}
\alsoaffiliation[Center for Nanostructured Graphene, Technical University of Denmark, DK-2800 Kgs. Lyngby, Denmark]{Center for Nanostructured Graphene}
\author{Martijn~Wubs}
\affiliation[Department of Photonics Engineering, Technical University of Denmark, DK-2800 Kgs. Lyngby, Denmark]{DTU Fotonik}
\alsoaffiliation[Center for Nanostructured Graphene, Technical University of Denmark, DK-2800 Kgs. Lyngby, Denmark]{Center for Nanostructured Graphene}
\author{N.~Asger~Mortensen}
\affiliation[Department of Photonics Engineering, Technical University of Denmark, DK-2800 Kgs. Lyngby, Denmark]{DTU Fotonik}
\alsoaffiliation[Center for Nanostructured Graphene, Technical University of Denmark, DK-2800 Kgs. Lyngby, Denmark]{Center for Nanostructured Graphene}
\email{asger@mailaps.org}
\title{Plasmonic eigenmodes in individual and bow-tie graphene nanotriangles}


\begin{document}
\begin{abstract}
  Serving as a new two-dimensional plasmonic material, graphene has stimulated an intensive study of its optical properties which benefit from the unique electronic band structure of the underlying honeycomb lattice of carbon atoms. In classical electrodynamics, nanostructured graphene is commonly modeled by the computationally demanding problem of a three-dimensional conducting film of atomic-scale thickness. Here, we propose an efficient alternative two-dimensional electrostatic approach where all the calculation procedures are restricted to the plane of the graphene sheet. To explore possible quantum effects, we perform tight-binding calculations, adopting a random-phase approximation. We investigate the multiple plasmon modes in triangles of graphene, treating the optical response classically as well as quantum mechanically in the case of both armchair and zigzag edge termination of the underlying atomic lattice. Compared to the classical plasmonic spectrum which is "blind" to the edge termination, we find that the quantum plasmon frequencies exhibit blueshifts in the case of armchair edge termination, while redshifts are found for zigzag edges. Furthermore, we find spectral features in the zigzag case which are associated with electronic edge states not present for armchair termination. Merging pairs of such triangles into dimers, the plasmon hybridization leads to energy splitting in accordance with plasmon-hybridization theory, with a lower energy for the antisymmetric modes and a smaller splitting for modes with less confinement to the gap region. The hybridization appears strongest in classical calculations while the splitting is lower for armchair edges and even more reduced for zigzag edges. Our various results illustrate a surprising phenomenon: Even 20 nm large graphene structures clearly exhibit quantum plasmonic features due to atomic-scale details in the edge termination.
\end{abstract}

\section{Introduction}

The collective excitations of conduction electrons in noble metals have been of great interest for a very long time. These excitations known as
plasmons play an important role in the optical properties of metals. Through strong plasmon-photon interactions, metals can support important phenomena, such as focusing beyond the diffraction limit,\cite{npho_4_83} squeezing the light down to nanoscale,\cite{nmat_9_205} and large local field enhancement.\cite{nature_424_824} Due to these features, plasmons in metals give rise to various potential applications, and especially form a bridge between the worlds of photonics and electronics which commonly work at different length scales.\cite{science_311_189} Developments in nanofabrication technology have stimulated a series of plasmon-based devices like waveguides,\cite{npho_2_496} filters,\cite{nl_12_4349} switches,\cite{nl_12_4977} and modulators.\cite{nl_9_4403} In many respects, plasmonic devices open a door to a better performance in speed and size, holding potential for faster dynamics than electronic devices while still having a smaller size footprint than the common all-dielectric photonic devices. However, the inherent Joule loss in metals severely hampers many practical applications of plasmonics.\cite{nmat_11_573} Alternatively, attempts have already been made to study plasmonics in materials other than metals,\cite{lpr_4_795} for example doped semiconductors\cite{pssrrl_4_295} and superconductors.\cite{apl_97_111106, npho_6_259}

Here we study the plasmonic properties of graphene flakes. In its pristine form graphene is a semimetal, but with appropriate doping it is emerging as a promising plasmonic material as well.\cite{prb_75_205418, prb_80_245435, nl_11_3370, ijmpb_27_1341001,Abajo:2014}
The charge carriers in graphene obey linear energy dispersion at lower energies close to the Dirac points, thus resembling the linear dispersion of photons.\cite{nmat_6_183, rmp_81_109, rmp_83_407} Experimental investigations of carrier transport show that the mobility limited by impurity scattering can exceed 15.000~$\textrm{cm}^2/\textrm{V}\textrm{s}$ at room temperature,\cite{nmat_6_183} which gives the intrinsic loss in graphene one order of magnitude less than the noble metals. Despite relaxation due to phonon scattering,\cite{npho_7_394, nl_14_2907} graphene achieves superior plasmonic performance in propagation length and field enhancement.\cite{nature_438_197, nature_438_201} The carrier density in graphene may be adjusted by electrostatic gating, which results in actively tunable plasmons beyond structural variations in metals, as has already been demonstrated experimentally.\cite{nnan_6_630, acsnano_7_2388} With the typical doping levels, the plasmonic response is generally in the terahertz (THz) to mid-infrared frequency range, thus allowing new progress in THz technology.\cite{acsnano_8_1086}

Because of these attractive plasmonic properties, it is worth to comprehensively study the optical properties of graphene. Here the fundamental quantity is the dielectric function. For graphene systems, the dielectric function can be obtained within the framework of linear-response theory and the random-phase approximation (RPA).\cite{prb_75_205418, prb_80_245435, njp_8_318} For infinite graphene sheets, the derived two-dimensional (2D) dielectric function $\varepsilon(q,\omega)$ is a function of both frequency and momentum. This is different from common three-dimensional (3D) photonic materials which are usually well-described by frequency-dependent functions, while spatial dispersion is negligible for good dielectrics and most metals (beyond the nanoscale). Two common approximations in the modelling of graphene structures are to adopt the local-response approximation (applying the small-$q$ limit) and to model graphene as a very thin conducting film, yet preserving its 3D representation.\cite{acsnano_6_431, prl_108_047401} Using dielectric functions so obtained, one can solve Maxwell's equations for arbitrarily shaped flakes of nanostructured graphene. For very small flakes of characteristic dimension $R$ ($R\sim \lambda_F$ with $\lambda_F\sim 10$\,nm being the Fermi wavelength corresponding to a Fermi energy of $\mu=0.4$\,eV), the common assumption $qR\gg 1$ is jeopardized and nonlocal response turns important. In this regime, both semiclassical hydrodynamic\cite{prb_87_195424, arxiv_1407_3920} and full quantum approaches have been proposed,\cite{acsnano_6_1766, prl_110_187401}  similar to those recently developed for metals.\cite{ncom_5_3809, ncom_5_3548}. While previous studies have mainly focused on the optically bright dipole mode, here we will illustrate that structured graphene is also rich on higher-order modes. Although the latter are typically not excited by far-field radiation, they may be probed by near-field optical spectroscopy and/or electron energy loss spectroscopy (EELS).

In this article, we study plasmon properties in individual graphene nanostructures and in dimers of such structures by means of both classical and quantum methods. In particular, we consider triangles of graphene and bow-tie structures formed by such triangles, while our methods can also be applied to other geometries (as we show in the Supplemental Material).

In our classical electrodynamical considerations, we treat the nanostructures as 2D materials characterized by a smooth surface conductivity (employing the sheet conductivity derived for bulk graphene), and formulate a closed-form eigenvalue problem on a 2D domain. Numerical solutions in arbitrarily shaped geometries are enabled by finite-element calculations. By its nature, this classical approach neglects the atomic details of the graphene flake. Some aspects e.g. of zigzag termination can be effectively accounted for by additional conductive channels\cite{arxiv_1407_3920}, but we will not adopt such effective schemes in our classical calculations here.

In our quantum treatment, we employ a tight-binding description\cite{acsnano_6_1766, prl_110_187401} to account for the actual position of all atoms in the flake and in particular the edge atoms which have the possibility for either armchair or zigzag configurations (More configurations can arise from the mixture of these two configurations, but they will not be discussed here). In both the classical and the quantum calculations, multiple plasmon modes are extracted including dipole, multipole, and breathing modes. Their hybridized counterparts in bow-tie nanostructures are also discussed. We show that plasmon excitations and hybridizations are extremely sensitive to the electronic edge effects. This illustrates how quantum plasmonics can manifest itself in graphene structures with dimensions much exceeding the length scales for nonlocal response in individual noble-metal nanoparticles.\cite{ncom_5_3809}

\section{Results and Discussion}

\subsection{Classical Description}
Modern computational electromagnetics is commonly optimized to explore the interaction of radiation with matter in a three-dimensional space, so that two-dimensional material problems are typically not efficiently addressed with existing numerical schemes. For example, a pragmatic approach is to simply mimic the atomically thin graphene layer with a homogenous dielectric film of a finite, yet small thickness $t$. This assumed 3D film has an effective bulk permittivity, $\varepsilon(\omega)=\varepsilon_0+i\sigma(\omega)/(\omega t)$ with $\sigma(\omega)$ denoting the surface conductivity as obtained from e.g.\ the local-response limit of the RPA.\cite{acsnano_6_431, prl_108_047401} Evidently, the effective thickness $t$ should be chosen sufficiently small compared with all other characteristic dimensions, yet sufficiently thick that meshing stays computationally feasible and the numerical problem remains tractable. Optimizing this tradeoff does not necessarily give an efficient method. Alternatively, in nanostructures with high symmetry, e.g.\ in ribbons\cite{prb_84_161407, prb_85_081405} or disks\cite{prb_33_5221,arxiv_1407_3920}, one may take advantage of modal expansion methods -- which, however, is not an appealing choice for more general structures. In the following, we develop a 2D finite-element approach to efficiently solve the electromagnetic problem self-consistently for graphene in terms of the electric potential and induced charge in general structural configurations.

With the typical sub-eV doping levels, plasmonic resonances typically occur in the mid-infrared regime. The associated free-space wavelength ( $\sim\!10\:\mu\mathrm{m}$) is then much larger than the geometrical extent of the hosting graphene nanostructures ($\sim\! 10-100\:\mathrm{nm}$). For such problems the electrostatic approximation is excellent. As a computationally very attractive consequence, the electric and constitutive response is governed by two coupled scalar equations for the potential $\phi$ and the induced density $\rho$. In particular, we note that the total potential $\phi(\bm r)$ is governed by Coulomb's law
\begin{subequations}
\begin{equation}\label{eq:coulombslaw}
\phi(\bm r) =\phi_{\rm ext}(\bm r)+\frac{1}{4\pi\varepsilon_{\rm s}L}\int_{2\textrm{D}}\mathrm{d}\bm r^\prime\frac{\rho(\bm r^\prime)}{|\bm r/L-\bm r^\prime/L|},
\end{equation}
where $\phi_{\rm ext}(\bm r)$ denotes the external potential, $L$ is an auxiliary quantity such as the feature length of the structure which make the surface integral dimensionless, $\rho(\bm r^\prime)$ the induced surface charge density, $\varepsilon_{\mathrm{s}}=(\varepsilon_{\rm above}+\varepsilon_{\rm below})/2$ the averaged dielectric constant of the medium above and below graphene. For simplicity, we only consider freely suspended graphene, so we will use $\varepsilon_{\mathrm{s}}=1$ throughout the remaining part of the paper.
The other scalar equation is obtained by inserting the constitutive equation $\bm J_{2\mathrm{D}} = -\sigma(\omega)\nabla_{2\textrm{D}}\phi(\bm r)$ into the continuity equation $i\omega\rho(\bm r)=\nabla\cdot\bm J_{2\mathrm{D}}$, which for $\bm r$ restricted to the plane of the graphene structure gives
\begin{equation}\label{eq:constitutiveEQ}
\rho(\bm r) =\frac{i\sigma(\omega)}{\omega}\nabla_{2\textrm{D}}^2\phi(\bm r),
\end{equation}
\end{subequations}
with $\nabla_{2\mathrm{D}}^2$ the 2D Laplace operator. Equation~\eqref{eq:constitutiveEQ} is solved subject to the assumption of charge neutrality, i.e.\ $\int_{2\mathrm{D}}\mathrm{d}\bm r\: \rho(\bm r)=0$, implying that $\hat{\bm n}\cdot \nabla_{2\mathrm{D}}\phi(\bm r)=0$ on the boundary of the domain, with $\hat{\bm n}$ denoting the in-plane surface normal.
The density $\rho$ in \eqref{eq:constitutiveEQ} is restricted to the graphene plane.
It may be obtained from a closed-form equation  by eliminating the potential in \eqref{eq:constitutiveEQ} with the help of \eqref{eq:coulombslaw} (see Methods for additional details).\cite{prb_86_125450}. Once $\rho$ within the graphene plane is thus obtained, the potential $\phi$ in the entire space can be evaluated  via~\eqref{eq:coulombslaw}.


Within the framework of the finite-element method (FEM), both equations~(\ref{eq:coulombslaw}) and (\ref{eq:constitutiveEQ}) can be recast as matrix equations. Concretely, by denoting the FEM-discretized potentials and induced charge densities by vectors, we find the equations $\bm \phi= \bm \phi_{\rm ext}+(4\pi\varepsilon_sL)^{-1}\cdot\mathbf{A}\bm\rho$ and $\bm \rho =i\sigma(\omega)\omega^{-1}\cdot\mathbf{B}\bm\phi $, which we combine to get
\begin{equation}\label{eq:matrixform}
\left[\mathbf{1}-f(\omega)\mathbf{B}\mathbf{A}\right]\bm\rho
=i\sigma(\omega)\omega^{-1}\cdot\mathbf{B}\bm\phi_{\rm ext},
\end{equation}
where $\mathbf{A}$ and $\mathbf{B}$ are geometry-dependent square matrices representing the Coulomb integral in Eq.~(1a) and the Laplacian in the Poisson equation~(1b) [see Methods part below for more details], while $f(\omega)=i\sigma(\omega)/(4\pi\varepsilon_sL\omega)$ is a geometry-independent scalar.\cite{acsnano_7_2388} Finally, the matrix in the square brackets on the left-hand side of Eq.~(\ref{eq:matrixform}) represents the effective frequency-dependent dielectric function $\varepsilon^{\textsc{cla}}(\omega)$. In the absence of an external potential ($\bm\phi_{\rm ext}=0$), Eq.~\eqref{eq:matrixform} becomes an eigenvalue problem for the matrix $\mathbf{B}\mathbf{A}$. The resulting eigenvalues $\lambda_n$ are associated with plasmon frequencies $\omega_n$ through $f(\omega_{n})=\lambda_{n}^{-1}$, and the associated eigenvectors are  induced charge densities ${\bm\rho}_{n}$ in a finite-element representation. The corresponding eigenpotentials are denoted as ${\bm\phi}_{n}$, and within the graphene plane they can be computed directly as ${\bm\phi}_{n}=\mathbf{A}{\bm \rho}_n$. Following this classical approach, all plasmonic eigenmodes for a specific structure can be obtained as the solution of a single eigenvalue problem. This constitutes an attractive computational approach that can give direct insight in the classical plasmonic eigenstates that one would be able to probe with various experimental techniques.

\subsection{Quantum Mechanical Tight-Binding Description}
In a quantum mechanical formalism, there are two key computational components: (i) electronic band structure, and (ii) determination of response functions. The graphene $\pi$ and $\pi^*$ bands (valence and conduction bands respectively) originating from the carbon $p_z$ orbitals are well separated in energy from the four $\sigma$ bands arising from $sp^2$ hybridization. The dynamics of low-energy excitations in graphene are well-described by inclusion of just the $\pi$ bands, which can be determined by a simple tight-binding model in a nearest-neighbor approximation.\cite{pr_71_622, prb_66_035412}
Specifically, a graphene nanostructure with $N$ carbon atoms results in an $N\times N$ matrix representation of the tight-binding Hamiltonian with elements determined by the $p_z$ orbital hopping integral. A direct diagonalization of the Hamiltonian yields $N$ eigenvalues and eigenvectors, corresponding to the electronic energy levels and the wave functions, respectively. The non-interacting density response function, or polarizability matrix $\mathbf{\chi^0}(\omega)$ is then built from the electronic states whose elements are given by \cite{prb_75_205418, prb_80_245435, njp_8_318}
\begin{equation}\label{eq:rpa}
\chi^0_{ll^\prime}(\omega) = 2\sum_{jj^\prime}(f_j-f_{j^\prime})\frac{\psi_{j^\prime l}^*\psi_{jl}\psi_{jl^\prime}^*\psi_{j^\prime l^\prime}} {\epsilon_j-\epsilon_{j^\prime}-\hbar(\omega+i\tau^{-1})},
\end{equation}
where $f_j=1/[\text{exp}((\epsilon_j-\mu)/k_{\textsc{b}}T)+1]$ denotes the Fermi--Dirac distribution function associated with the state with energy $\epsilon_j$ and wave function $\psi_{jl}$ ($l$ the labels of the carbon atoms), while the factor 2 accounts for spin degeneracy. In both classical (or called semi-classical due to the conductivity including Fermi--Dirac distribution function) and quantum calculations, states are populated in accordance with a Fermi energy of $\mu=0.4\:\text{eV}$ and a temperature $T=300\:\text{K}$. We phenomenologically account for relaxation losses through $\hbar\tau^{-1}=6\:\mathrm{meV}$, commensurate with experimental data at the considered doping level\cite{science_341_620}. We use an efficient method to compute the non-interacting density response matrix $\mathbf{\chi^0}(\omega)$, based on Hilbert and fast Fourier transforms (see Ref.~\citenum{acsnano_6_1766} and Methods section below).

Including the effects of a self-consistent Hartree interaction, i.e. within the RPA, the interacting polarizability is given by~\cite{njp_8_318}
\begin{equation}
\mathbf{\chi^{\textsc{rpa}}}(\omega)=\frac{1}{1-\mathbf{V}\mathbf{\chi^0}(\omega)}\cdot\mathbf{\chi^0}(\omega),
\end{equation}
with the Coulomb interaction $V_{ll^\prime}\propto1/|\bm r_l -\bm r_{l^\prime}|$ for $l\neq l^\prime$, and a self-interaction of 0.58 atomic units at $l=l^\prime$. \cite{acsnano_6_1766} The poles of $\mathbf{\chi^{\textsc{rpa}}}(\omega)$ or equivalently the zeros of the denominator
\begin{equation}
\mathbf{\varepsilon^{\textsc{rpa}}}(\omega) = \mathbf{1}-\mathbf{V}\mathbf{\chi^0}(\omega)
\end{equation}
give the plasmon frequencies. Since $\mathbf{\varepsilon^{\textsc{rpa}}}(\omega)$ is a matrix, we follow Ref.~\citenum{prb_86_245129} and look for the eigenvalues $\varepsilon_n(\omega)$ of the matrix which are approaching zero. In practice there is also loss, for example due to $\hbar\tau^{-1}$ in $\mathbf{\chi^0}(\omega)$. For real-valued frequencies the  $\varepsilon_n(\omega)$ are therefore complex-valued, with imaginary parts denoting the plasmon peak broadening. On the real frequency axis it is therefore more accurate to define plasmon frequencies from the local maximum of $-\text{Im}[\varepsilon_n^{-1}(\omega)]$.\cite{prb_86_245129, prb_87_235433}

Numerically, the eigenvalues $\varepsilon_n(\omega)$ are obtained by diagonalizing the RPA dielectric function $\varepsilon^{\textsc{rpa}}(\omega)$ for each frequency. An $N$-atom nanostructure entails $N$ distinct eigenvalues. Out of these we focus in the following on eigenvalues with largest and second-largest value of $-\text{Im}[\varepsilon_n^{-1}(\omega)]$. Their corresponding eigenvectors are the induced charge densities $\bm{\rho}_n$, and similarly the eigenpotentials $\bm{\phi}_n$ can be obtained by performing coulomb integral. For comparison with the quantum treatment, we also calculate the eigenvalue loss spectrum in the classical framework by carrying out diagonalization of the classical effective dielectric function $\varepsilon^{\textsc{cla}}(\omega)$.

\subsection{Plasmonic Eigenmodes in Individual Triangles}
\begin{figure}
\includegraphics[width=12.0cm]{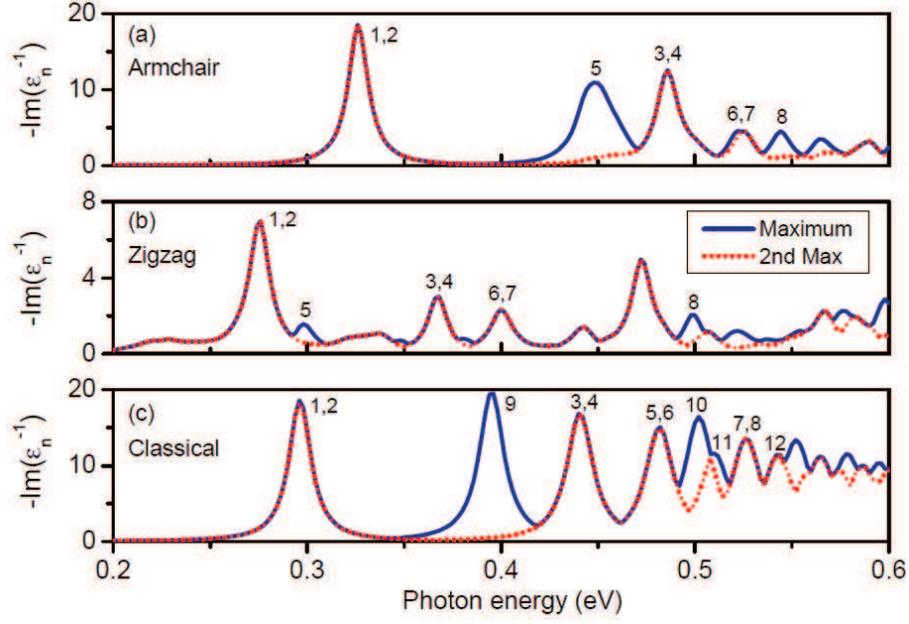}
\caption{The eigenvalue loss spectrum $-\text{Im}[\varepsilon_{n}^{-1}(\omega)]$ in equilateral graphene triangles of sidelength $20\:\text{nm}$. Each peak defines a plasmon mode (labeled by $n=1,2,3,\ldots$ in order of decreasing imaginary part), and the coincidence of the maximum (blue solid) and the second maximum (red dotted) indicates the energy degeneracy. Results of the quantum tight-binding method in (a) for armchair edges, and in (b) for zigzag edges, while classical results are given in (c).}
\label{fig:1}
\end{figure}

\begin{figure}
\includegraphics[width=12.0cm]{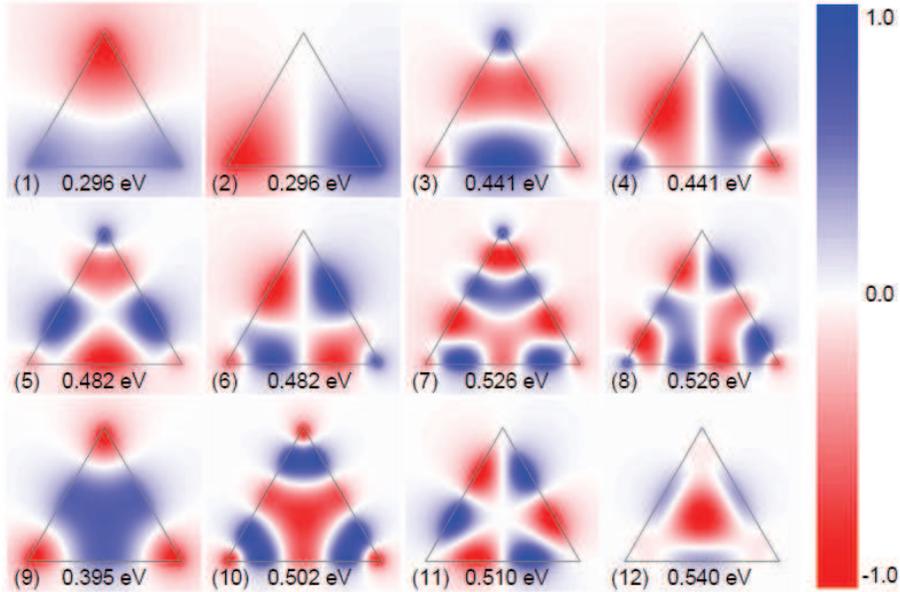}
\caption{In-plane potentials ${\bm\phi}_{n}$ for the  twelve lowest-energy plasmon modes calculated in the classical approach, from the eigenvectors of the matrix pair $\mathbf{B}\mathbf{A}$ of Eq.~(\ref{eq:matrixform}).}
\label{fig:2}
\end{figure}

The calculated eigenvalue loss spectrum for 20~nm graphene equilateral triangles is shown in Figure~\ref{fig:1}. In the quantum description we distinguish between zigzag and armchair edge terminations, see Supporting Information. Multiple plasmon peaks are visible in the considered frequency regime. Additionally, at several frequencies, the two considered loss functions (largest and second largest value of $-\text{Im}[\varepsilon_n^{-1}(\omega)]$) exhibit nearly identical values, while at other frequencies one can be resonant while the other one is not. This is in full accordance with group-theoretical considerations for our structure with $m$-fold rotational symmetry where the $C_{m}$ point group leads to either non-degenerate eigenstates or pairs of eigenstates with a double degeneracy\cite{Sakoda:2005}. The degeneracy can be explored further by considering the eigenmodes, expressed e.g. by the in-plane potential, and in particular their symmetries. In the classical approach, the eigenmodes appear as eigenvectors of the matrix $\mathbf{B}\mathbf{A}$ of Eq.~(\ref{eq:matrixform}).
Considering the two lowest eigenstates causing the resonance around 0.3\,eV in Figure~\ref{fig:1}(c), we numerically find the eigenfrequencies to be 0.2964~eV and 0.2963~eV. The small energy difference of 0.1~meV illustrates the numerical accuracy (symmetry breaking) associated with the fact that our finite-element mesh does not comply with the threefold rotational symmetry of the graphene triangle. In Figure~\ref{fig:2} we show corresponding in-plane potential distributions of the twelve lowest-energy  eigenmodes, again calculated in the classical framework. The eigenmodes are responsible for the primary features of Figure~\ref{fig:1}c; specifically, the loss-function exhibits peaks at the resonance energies of the eigenmodes. The peaks are each assigned a label ($n=1,2,3,\ldots$), corresponding to the eigenmode enumeration in Figure~\ref{fig:2}. A one-to-one correspondence is evident and whenever the spectrum in Figure~\ref{fig:1} suggests a pair of degenerate states, the corresponding modes in Figure~\ref{fig:2} support that they are indeed pairs of orthogonal and degenerate states. The energy degeneracies exhibited here are a direct consequence of the symmetries of the considered nanostructure, as required by group theory.\cite{jpcc_116_14591}

The plasmon modes 1 through 8, being doubly degenerate, are either symmetric or antisymmetric with respect to the mirror symmetry plane. The dipole modes, 1 and 2, with the electric field being polarized orthogonal to each other, are of particular interest due to their strong coupling to optical fields. They can be excited directly by far-field techniques, and the plasmonic local field enhancement is concentrated at the vertices. The modes 3 through 8 penetrate significantly into the bulk, and can be considered as hybridized modes originating from interaction between dipole and bulk modes, because the patterns at the vertices are similar to dipole modes 1 and 2; in addition, the modes $3 - 6$ have very small net dipole momenta, and can couple to far-field radiation. The modes $9 - 12$ are no longer doubly degenerate, and exhibit threefold rotational symmetry around the center. Although optically dark, these modes are still detectable by suitable near-field techniques. As an example, in an EELS experiment the breathing mode 12 would exhibit the strongest coupling to a nanometer-sized electron beam if this beam were passing through the center of the graphene triangle.\cite{nl_12_5780}
\begin{figure}[h]
\includegraphics[width=12.0cm]{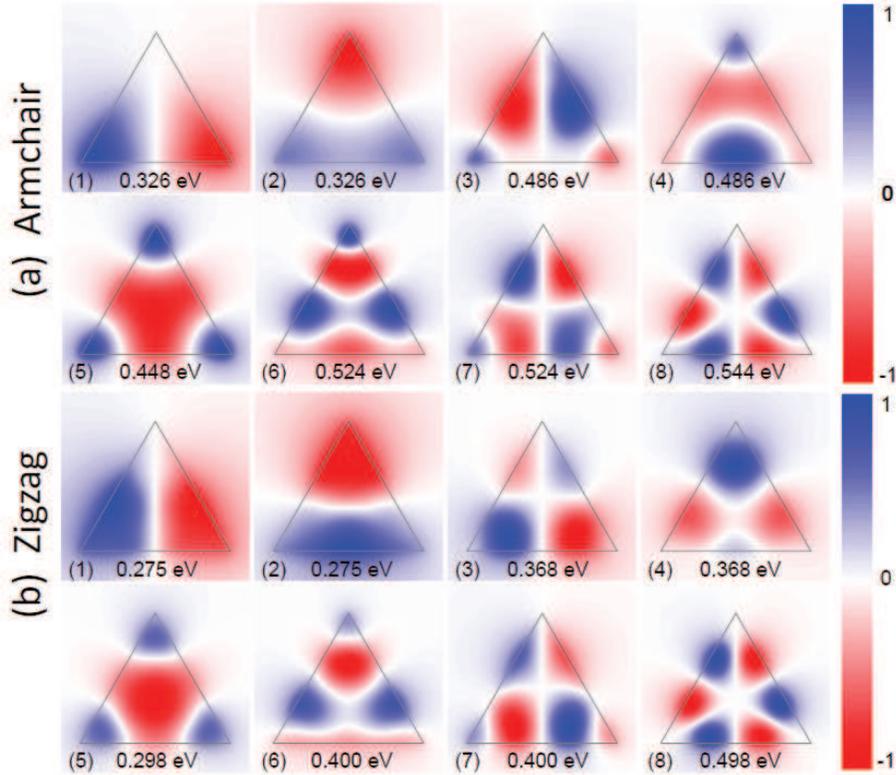}
\caption{In-plane potential $\bm{\phi}_n$ extracted from the eigenvalue loss spectrum calculations. (a) The lowest 8 plasmon modes in an armchair triangle; (b) the corresponding plasmon modes in a zigzag triangle. }
\label{fig:3}
\end{figure}

Having described our classical results for graphene triangles, let us now turn to our corresponding tight-binding quantum results. In the quantum description, we calculate the eigenvalue loss spectrum, identify the plasmon mode eigenfrequencies, and then extract the corresponding eigenmodes.
Due to the geometrical symmetry, the plasmon eigenmodes should exhibit the same energy degeneracy features as the equilateral triangles in classical calculations, for instance in Figure panels~\ref{fig:1}(a) and ~\ref{fig:1}(b) several doubly degenerate plasmon modes occur. Figure~\ref{fig:3} shows the wave patterns from the quantum calculations, corresponding to the peak labeling in Figure~\ref{fig:1}(a) and \ref{fig:1}(b). We observe that for the armchair case the modes of the same type are blueshifted when compared to their classical counterparts. On the contrary, zigzag termination cause lower plasmon energies with a net redshift compared to the classical case. As an example, the eigenfrequencies of dipole modes are $0.326\:\text{eV}$, $0.275\:\text{eV}$, and $0.296\:\text{eV}$ for the armchair, zigzag, and classical cases, respectively. The associated mode patterns are only slightly different, yet it is clearly seen from the dipole modes, that in zigzag-terminated triangles the mode spreads much more into the bulk while for armchair termination the mode concentrates at the vertices in the same manner as for the classical results. This trend becomes even more evident in the modes 3 and 4 of which the patterns show no hot spots at the vertices. The somewhat different and unusual mode behavior for zigzag-terminated triangles is due to the electronic edge states which do not occur for armchair termination (see Supplementary Figure S3 for additional details). Similar edge-state effects on plasmon excitations have been discussed for graphene ribbons\cite{acsnano_6_1766} and disks\cite{prl_110_187401} and recently their importance has been illustrated explicitly through analytical calculations.\cite{arxiv_1407_3920}

\subsection{Plasmon Hybridization in Bow-Tie Triangles}
\begin{figure}[h]
\includegraphics[width=12.0cm]{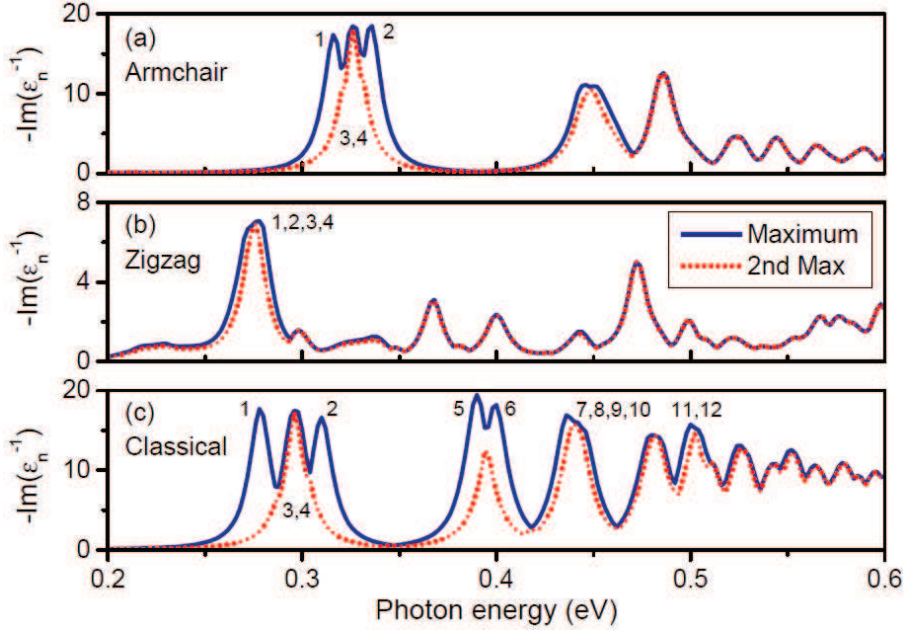}
\caption{The eigenvalue loss spectrum $-\text{Im}\varepsilon_n(\omega)^{-1}$ in graphene bow-tie triangles with gap width $0.5\:\text{nm}$. Results obtained from quantum calculations in armchair triangles (a) and in zigzag triangles (b) are compared with classical calculations in (c).}
\label{fig:4}
\end{figure}
\begin{figure}[h]
\includegraphics[width=12.0cm]{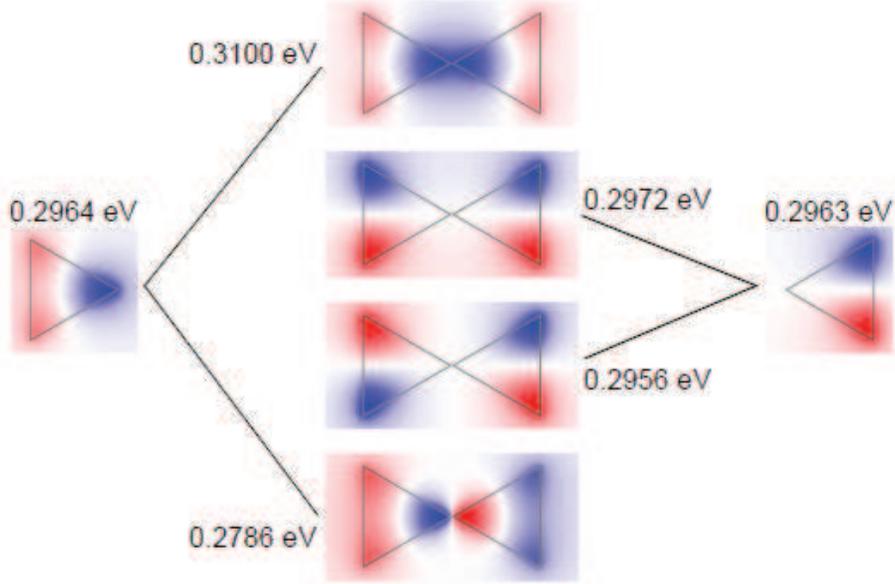}
\caption{A schematic diagram of the dipole mode hybridization in classical calculations. There is a larger energy splitting for $x$-polarized dipole, and the antisymmetrically coupled modes have lower energy for both polarizations. Here the gap distance is 0.5 nm.}
\label{fig:5}
\end{figure}

Plasmon hybridization is of both fundamental and practical importance.\cite{science_302_419, nl_4_899} Hybridization through tuning of the gap distance can be used achieve better performance through careful design, such as the field enhancement in dimers\cite{nl_9_887} and the sensing capabilities in Fano structures.\cite{nl_8_3983}
Here, we study the plasmon hybridization in graphene bow-tie triangles, using the same classical and quantum methods as for individual triangles above. Figure~\ref{fig:4} shows the calculated eigenvalue loss spectra for a gap width of $0.5\:\text{nm}$. There are four modes ($n=1,2,3,4$) in the classical calculations, originating from the four (accounting degeneracy) low-energy dipole modes of the two un-coupled triangles. The hybridization process is illustrated in Figure~\ref{fig:5} with a focus on dipole modes, where energies are given with higher precision in order to display the tiny energy shifts associated with the hybridization. We find that each dipole mode in the individual triangles will split into two modes in the bow-tie triangles forming either bonding or antibonding states. The $x$-polarized dipole ($0.2964\:\text{eV}$, dipole aligned parallel to bow-tie axis) exhibits large energy splitting, and the corresponding bonding (antisymmetrically coupled) mode has lower energy. However, for the $y$-polarized dipole ($0.2963\:\text{eV}$, dipole aligned perpendicular to bow-tie axis) the reduced mode-overlap causes a very small energy splitting. In both cases, the bonding modes are optically active with a net dipole polarization along $x$ and $y$ direction, respectively.

We find a very similar behavior in the armchair-terminated bow-tie triangles shown in Figure~\ref{fig:4}a, but with smaller energy splitting, which originates from a weaker mode overlap and weaker coupling strength when compared to the classical calculations. In the zigzag-terminated bow-tie triangles (see Figure~\ref{fig:4}b), the coupling strength is even weaker and the $x$-polarized dipole exhibits no appreciable energy splitting when compared to the line width of the uncoupled resonances. As a result of this approximate degeneracy, the coupled system exhibits a single broad peak with all four modes merged together. In contrast to the dipole modes, the higher-order plasmon modes show a weak lifting of degeneracy for antisymmetrical and symmetrical states. We mention that the hybridization picture given in Figure~\ref{fig:5} is very general, also being satisfied in quantum calculations but with different eigenfrequencies (hybridization diagrams not shown).

\begin{figure}[h]
\includegraphics[width=8.0cm]{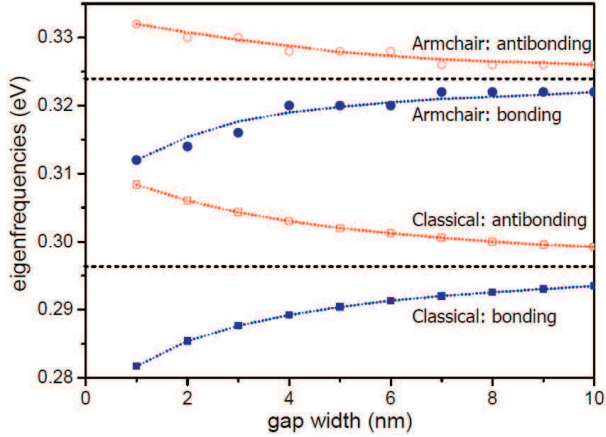}
\caption{The eigenfrequencies of the hybridized modes as a function of gap width for $x$-polarized dipoles in classical calculations and armchair triangles, respectively. The two dotted lines ($0.296\:\text{eV}$ and $0.324\:\text{eV}$) are the dipole eigenfrequencies associated with the individual triangles.}
\label{fig:6}
\end{figure}
The energy splitting or coupling strength depends on the gap width of the bow-tie structures, which can be investigated in the hybridization of $x$-polarized dipoles. We calculate the eigenfrequencies of the hybridized plasmon modes as a function of the width gap, and show the results in Figure~\ref{fig:6}. The modes in zigzag triangles exhibit very small energy splitting, so we do not show them here. Both in the classical calculations and armchair-terminated triangles, the energy splitting decreases as the gap width increases. The decrease is most pronounced for gap widths below $4\:\text{nm}$, while the variation is weaker for larger separations.


\begin{figure}[h]
\includegraphics[width=12.0cm]{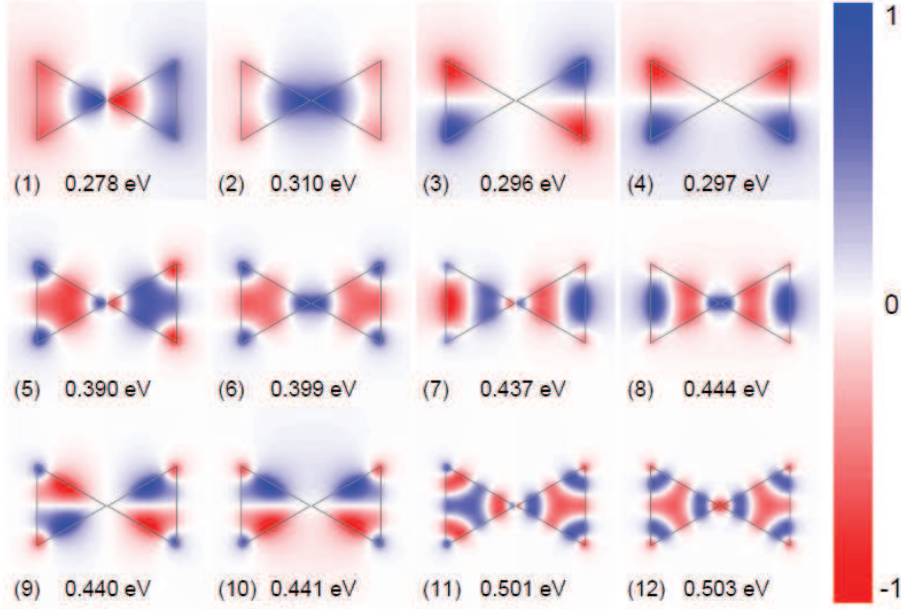}
\caption{Wave patterns for selected twelve plasmon modes calculated from the eigenvectors of the matrix pair $\mathbf{B}\mathbf{A}$ of Eq.~(\ref{eq:matrixform}).}
\label{fig:7}
\end{figure}

We note that the hybridization of other dimer plasmon modes (other than dipole modes) can be analyzed with a similar result. Generally speaking, the eigenfrequencies of the resulting hybridized modes are decided by two factors: symmetry and coupling strength. Specifically, the antisymmetrically coupled modes (no matter which polarization) have lower energy and modes with less field concentration at the gap region cause weaker coupling and consequently exhibit smaller energy splitting. As a further evidence for this qualitative characterization, we show in Figure~\ref{fig:7} the selected twelve plasmon modes from classical calculations, corresponding to the peaks shown in Figure~\ref{fig:4}c. As compared with Figure~\ref{fig:2}, they can be understood as linear combinations of the wave patterns in individual structures. Likewise, it is straightforward to envision the wave patterns in armchair and zigzag bow-tie triangles based on the uncoupled modes from Figure~\ref{fig:3}.

\section{Conclusions}
In this article, we have considered and compared classical and quantum aspects of plasmonic eigenmodes in graphene triangular nanostructures. The 2D FEM-approach for calculation of the classical electromagnetic response represents a numerically highly efficient method for electrodynamics in general 2D morphologies of graphene structures in the electrostatic limit (see Supplementary Information for the calculation in hexagonal structures). The simple eigenvalue approach offers a direct pathway to extraction of all plasmonic eigenmodes, not limited to just the optically active, but including also dark modes and highly symmetric breathing modes. The quantum method adopted here is useful for investigating the quantum effects in plasmon excitations of smaller graphene structures, and it offers additional insight into the importance of the particular edge-termination of the underlying atomic lattice. By a sweep of the excitation energy, our calculation of the eigenvalue loss spectra enables direct identification of all plasmonic modes also in the quantum treatment.

We have applied both methods to equilateral triangles, of $20\:\mathrm{nm}$ sidelength, both in isolated and in bow-tie configurations. For the isolated nanotriangle we find that the plasmonic response of armchair-terminated triangles is qualitatively similar to the classical case, albeit with a significant and consistent blueshift of all resonances due to nonlocal response. Conversely, the response of zigzag-terminated triangles exhibits several significant differences from its classical counterpart. As a consequence of the existence of localized electronic edge states near zigzag edges, the eigenmodes extend further into the bulk, and are less intense at the vertices. Additionally, we observe a redshift and an pronounced readjustment of the loss-function intensity relative to the classical case.

In the bow-tie configuration we observe plasmon hybridization and associated eigenmode energy splitting, of varying degree depending on treatment; the largest splitting is observed in the classical approach, and the smallest in zigzag structures. Nevertheless, the effects of hybridization are qualitatively similar across the considered cases, with the antisymmetric hybridized modes exhibiting a lowered energy, and with the coupling strength - and associated energy splitting - decreasing when the constituent eigenmodes exhibit lower field intensities in the gap region.

\section{Methods}
\subsection{Classical Calculations}
The classical calculations are performed on the 2D graphene surface. We generate the triangular meshes within the graphene domain (see Supplementary materials for details), and approximate the integration of Eq.~\eqref{eq:coulombslaw} by summing all the elements. For example, the $j$th element has three vertices $l$, $m$, and $n$, and the area is $s_j$, and thus Eq.~(1a) becomes
\begin{equation}
\phi(\bm r) =\sum_{j}\phi_{\rm ext}\left(\bm r_c\right)+\frac{1}{4\pi\varepsilon_{\rm s}L}\sum_{j}s_j\frac{\rho(\bm r_c)}{|\bm r/L-\bm r_c/L|},
\end{equation}
where
\begin{subequations}
\begin{align}
&\bm r_c = \frac{\bm r_l+\bm r_m+\bm r_n}{3}, \\
&\rho(\bm r_c) = \frac{\rho_l+\rho_m+\rho_n}{3}.
\end{align}
\end{subequations}
We can obtain the matrix $\mathbf{A}$ from Eq.~(6). Instead of assembling matrices $\mathbf{B}$ from Eq.~\eqref{eq:constitutiveEQ} directly, we use an idea behind FEM and  multiply by $\phi_j^*(\bm r)$ and perform the integral, giving
\begin{equation}
\begin{split}
\int_{2\mathrm{D}} \! \mathrm{d}\bm r\:\phi_j^*(\bm r)\rho(\bm r) &=-\frac{\sigma(\omega)}{i\omega}\int_{2\mathrm{D}} \! \mathrm{d}\bm r\:\phi_j^*(\bm r)\nabla^2\phi(\bm r) \\
&=\frac{\sigma(\omega)}{i\omega}\int_{2\mathrm{D}} \! \mathrm{d}\bm r\:\nabla\phi_j^*(\bm r)\cdot\nabla\phi(\bm r),
\end{split}
\end{equation}
where the boundary condition $\hat{n}\cdot \nabla_{2\mathrm{D}}\phi(\bm r)=0$ has been applied at the second equality sign, and where $\phi_j(\bm r)$ is a linear test function at the $j$th element with $j$ running over all elements. Within a local coordinate system $(\eta, \xi)$, the position and wave function can be expressed as
\begin{subequations}
\begin{align}
&\bm r = (1-\eta-\xi)\bm r_l+\eta\bm r_m+\xi\bm r_n, \\
&\phi(\bm r) = (1-\eta-\xi)\phi_l+\eta\phi_m+\xi\phi_n,
\end{align}
\end{subequations}
and after straightforward algebra we obtain
\begin{equation}
\begin{split}
\nabla_{2D}\phi(\bm r) &= \left[(\phi_m -\phi_l)\frac{y_n-y_l}{2s_j}-(\phi_n-\phi_l)\frac{y_m-y_l}{2s_j}\right]\hat{e}_x \\
&+\left[-(\phi_m -\phi_l)\frac{x_n-x_l}{2s_j}+(\phi_n-\phi_l)\frac{x_m-x_l}{2s_j}\right]\hat{e}_y.
\end{split}
\end{equation}
Finally, in each element the left-hand side of Eq.~(8) yields
\begin{equation}
\mathbf{B}_L=\frac{s_j}{12}
\left[ \begin{array}{ccc}
2 & 1 & 1 \\
1 & 2 & 1 \\
1 & 1 & 2 \end{array} \right],
\end{equation}
and the right-hand side yields
\begin{equation}
\mathbf{B}_R =\frac{1}{4s_j}
\left[ \begin{array}{ccc}
|\bm r_m-\bm r_n|^2 & (\bm r_m -\bm r_n)\cdot(\bm r_n-\bm r_l) & (\bm r_m -\bm r_n)\cdot(\bm r_l-\bm r_m) \\
(\bm r_m -\bm r_n)\cdot(\bm r_n-\bm r_l) & |\bm r_n-\bm r_l|^2 & (\bm r_n -\bm r_l)\cdot(\bm r_l-\bm r_m) \\
(\bm r_m -\bm r_n)\cdot(\bm r_l-\bm r_m) & (\bm r_n -\bm r_l)\cdot(\bm r_l-\bm r_m) & |\bm r_l-\bm r_m|^2
\end{array} \right].
\end{equation}
The final matrix is given by $\mathbf{B} = \mathbf{B}_L^{-1}\mathbf{B}_R$. We model the bulk conductivity $\sigma(\omega)$ of graphene by its well-known local-response form \cite{jap_103_064302,epjb_56_281}
\begin{equation}
\begin{split}
\sigma(\omega)=&\frac{ie^2}{\pi\hbar}\frac{ k_{\textsc{b}}T}{\hbar(\omega+i\tau^{-1})}\bigg[\frac{\epsilon_{\textsc{f}}}{k_{\textsc{b}}T}
+2\ln\big(e^{-\epsilon_{\textsc{f}}/k_{\textsc{b}}T}+1\big)\bigg] \\
&+\frac{e^2}{4\hbar}\bigg[\theta(\hbar\omega-2\epsilon_{\textsc{f}})
+\frac{i}{\pi}\ln\bigg|\frac{\hbar\omega-2\epsilon_{\textsc{f}}}{\hbar\omega+2\epsilon_{\textsc{f}}}\bigg|\bigg],
\end{split}
\end{equation}
with the first and second terms due to intra- and interband dynamics, respectively.

\subsection{Quantum Calculations}
The tight-binding Hamiltonian for the $\pi$-electrons is constructed by considering only nearest-neighbor interactions with a hopping strength $t = 2.8\:\text{eV}$. The associated Hamiltonian matrix-representation is real-valued and symmetric, giving rise to real eigenvalues and eigenvectors.

The direct evaluation of the noninteracting density response matrix $\chi^0(\omega)$ of Eq.~\eqref{eq:rpa} requires significant computational resources and time, amounting to $\sim\!\! N^4$ operations, which must additionally be repeated for each distinct frequency. Significant reduction of computational complexity, to $\sim\!\! N^3$, can be achieved with the aid of Hilbert and fast Fourier transform (FFT), following a procedure developed in density-functional theory (DFT),\cite{prb_74_035101, prb_61_7172} and recently implemented in Ref.~\citenum{acsnano_6_1766} for the tight-binding model of graphene considered here. We adopt the same technique in our computations.

Furthermore, consideration of the symmetry of $\chi^0(\omega)$, i.e.$\chi^0_{ll^\prime}(\omega)=\chi^0_{l^\prime l}(\omega)$, leads to an additional reduction of the computational requirements.

\begin{acknowledgement}

We thank Wei Yan, Xiaolong Zhu, and Nicolas Stenger for stimulating discussions. The Center for Nanostructured Graphene is sponsored by the Danish National Research Foundation, Project DNRF58. This work was also supported by the Danish Council for Independent Research--Natural Sciences, Project 1323-00087.

\end{acknowledgement}

\section{References and Notes}
\mciteErrorOnUnknownfalse
\providecommand{\latin}[1]{#1}
\providecommand*\mcitethebibliography{\thebibliography}
\csname @ifundefined\endcsname{endmcitethebibliography}
  {\let\endmcitethebibliography\endthebibliography}{}

\end{document}